\documentclass[conference]{IEEEtran}

\IEEEoverridecommandlockouts
\usepackage{cite}
\usepackage{amsmath,amssymb,amsfonts}
\usepackage{algorithmic}
\usepackage{graphicx}
\usepackage{textcomp}
\usepackage{xcolor}
\usepackage{booktabs}
\usepackage{multirow}
\usepackage{float}
\usepackage{tabularx}
\usepackage{array}
\usepackage{adjustbox}
\usepackage{ltablex} 
\usepackage{longtable} 
\usepackage{tabularx} 
\usepackage{hyperref}
\usepackage{adjustbox}
\usepackage{lipsum} 
\usepackage{tabto}

\usepackage[utf8]{inputenc}
\usepackage{array}
\usepackage{soul}
\usepackage{comment}
\usepackage[most]{tcolorbox}
\usepackage[T1]{fontenc}
\usepackage{caption}
\usepackage{subcaption}
\usepackage{xspace}

\newcommand{\AlgoName}{\textsc{BarkPlug v.2}\xspace}

\def\BibTeX{{\rm B\kern-.05em{\sc i\kern-.025em b}\kern-.08em
    T\kern-.1667em\lower.7ex\hbox{E}\kern-.125emX}}
\begin{document}

\title{From Questions to Insightful Answers: Building an Informed Chatbot for University Resources
}

\author{
    \IEEEauthorblockN{
        Subash Neupane\IEEEauthorrefmark{1}, 
        Elias Hossain\IEEEauthorrefmark{1}, 
        Jason Keith\IEEEauthorrefmark{2},
        Himanshu Tripathi\IEEEauthorrefmark{1},\\
        Farbod Ghiasi\IEEEauthorrefmark{1},
        Noorbakhsh Amiri Golilarz\IEEEauthorrefmark{1}
        Amin Amirlatifi\IEEEauthorrefmark{2},
        Sudip Mittal\IEEEauthorrefmark{1},
        Shahram Rahimi\IEEEauthorrefmark{1}
    }
    \IEEEauthorblockA{
        \IEEEauthorrefmark{1}
        \textit{Department of Computer Science \& Engineering} \\
        \textit{Mississippi State University} \\
        \{sn922, mh3511, ht557, fg289\}@msstate.edu,\{amiri, mittal, rahimi\}@cse.msstate.edu
    }
    \IEEEauthorblockA{
        \IEEEauthorrefmark{2}
        \textit{Dave C. Swalm School of Chemical Engineering} \\
        \textit{Mississippi State University} \\
        \{keith, amin\}@che.msstate.edu
    }

}

\maketitle

\begin{abstract}

This paper presents \AlgoName, a Large Language Model (LLM)-based chatbot system built using Retrieval Augmented Generation (RAG) pipelines to enhance the user experience and access to information {within academic settings.} 
The objective of \AlgoName is to provide information to users about various campus resources, including academic departments, programs, campus facilities, and student resources at a university setting in an interactive fashion.  Our system leverages university data as an external data corpus and ingests it into our RAG pipelines for domain-specific question-answering tasks. We evaluate the effectiveness of our system in generating accurate and pertinent responses for Mississippi State University, as a case study, using quantitative measures, employing frameworks such as Retrieval Augmented Generation Assessment (RAGAS). Furthermore, we evaluate the usability of this system via subjective satisfaction surveys using the System Usability Scale (SUS). Our system demonstrates impressive quantitative performance, with a mean RAGAS score of 0.96, and satisfactory user experience, as validated by usability assessments.

\end{abstract}

\begin{IEEEkeywords}
Chatbot, LLM, RAG, University resources, information access
\end{IEEEkeywords}

\section{Introduction}

Colleges and universities invest significant time and resources into enhancing their websites to effectively communicate crucial information about the institution and available campus resources. The institutional website serves as its ``virtual face’’, the face it has chosen to present to the online world, including potential and current students, faculty, parents, alumni and general users \cite{meyer2011information}. Although these websites offer comprehensive information, they lack the capability to provide personalized responses to user queries. For instance, when a prospective student needs details about submitting ACT scores, wants to know their tuition and fees, or is unsure which parent's information to use on their FAFSA application, they must navigate through multiple webpages to find answers. This process frequently requires a considerable amount of time. Yet, at times, users' queries are left unanswered due to either unclear information or lack of personal interaction. 

Various campus resources and services, such as academic departments, career centers, admissions, registration, scholarships, and financial aid, are available to assist students with both academic and non-academic queries. These resources are equipped with dedicated officers who provide guidance to students. However, they are constrained by service-time limitation (may only be available during specific working hours) and may require an appointment, which might not always accommodate  busy student schedules. Additionally, delays in responses and longer wait times, particularly during application periods, can diminish prospective students' interest in the institution, ultimately affecting university revenue.

\begin{figure*}[htb]
\vspace{-5mm}
\centering
    \includegraphics[scale=0.70]{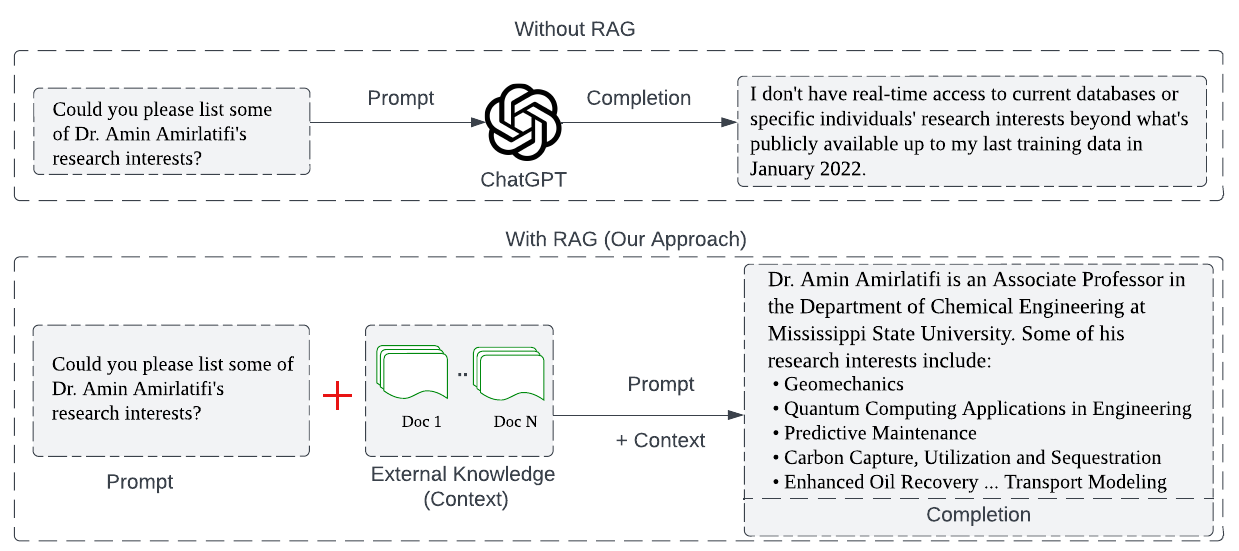}
    \caption{Comparative example of completion (response generation) without using the RAG approach versus using the RAG approach for a given user prompt related to specific individual at Mississippi State University.}
    \label{fig:overview}
    \vspace{-2mm}
\end{figure*}

To address these challenges, universities are currently employing conversational agents, also known as chatbots, to offer support to users. Chatbots are \emph{``software systems that mimic interactions with real people”}\cite{khatri2018alexa} by engaging in conversation through natural language using machine learning technology, specifically Natural Language Processing (NLP)\cite{bradevsko2012survey}. For instance, Arizona State University (ASU) developed a chatbot named \emph{Sunny} \cite{sunny} with the intentional design to offer emotional support to students, alongside providing information regarding ASU. Sunny efficiently addresses frequently asked questions such as inquiries about financial aid, academic advisors, and accessing ASU email accounts. Similarly, Georgia State University (GSU) introduced a virtual assistant named \emph{Pounce} \cite{pounce} to tackle obstacles to enrollment faced by students transitioning from high school to college. These obstacles encompassed tasks like financial aid applications, document submissions, immunization records, placement exams, and class registration. GSU reported a notable 22\% reduction in summer melt due to Pounce's assistance. Beyond admissions and enrollment, universities are increasingly deploying chatbots to aid students in their academic pursuits and campus life. One notable example is the chatbot \emph{Beacon} \cite{beacon} developed by Staffordshire University. Beacon offers personalized and responsive support, including information on timetables and answers to frequently asked questions. 

Apart from the higher education sector, chatbots are being increasingly adopted across a diverse range of industries and contexts including healthcare \cite{laranjo2018conversational, neupane2024medinsight}, cybersecurity\cite{franco2020secbot, mitra2024localintel}, retail\cite{chung2020chatbot}, and hospitality\cite{leung2020chatbot} among others due to their ability to emulate human conversations, automate services, and reduce human workload. The meteoric rise in interest in using chatbots by industries at present is attributed to the overwhelming success of ChatGPT.\footnote{https://chatgpt.com/}.  In fact, the global chatbot market size was valued at 5.39 billion dollars in 2023 which is expected to reach 42.83 billion dollars by 2033, according to a market research report \cite{globalchatbot} published by Spherical Insights \& Consulting. 

In this paper, we introduce \AlgoName - the second iteration of a chatbot system built for Mississippi State University (MSU), with an architecture that can be applied to any university setting. This system serves as an assistive tool, capable of leveraging all university resources to provide more intelligent analyses of university related content. It responds interactively, making access to relevant information easier for users. Compared to the other educational chatbots, \AlgoName is more comprehensive and covers several aspects of university functions and services. The development of our chatbot utilizes Retrieval Augmented Generation (RAG) \cite{lewis2020retrieval} techniques for response generation. RAG pipelines consist of two vital components: a retriever and a generator based on a Large Language Model (LLM). We opt for the RAG approach because pretrained LLMs, such as \emph{gpt-3.5-turbo}, alone cannot adequately answer domain-specific questions or perform well on data outside their training dataset, often resulting in hallucinated outputs. Figure \ref{fig:overview} provides a comparative overview of the response generation for a given user prompt without RAG and with RAG. As is evident from Figure \ref{fig:overview}, ChatGPT clearly fails to answer domain-specific questions, while \AlgoName, which uses the RAG approach, can accurately answer a user prompt. In our pipeline we utilize various campus resources such as information on \emph{academic departments, financial aid, admission, scholarships, dining, housing, and health center}
as a corpus of external data source for retrieval.

\AlgoName project's key contributions include: 

\begin{itemize}
  \item Design and development of a comprehensive chatbot system proficient in responding to a wide spectrum of queries pertaining to the diverse array of campus resources available at Mississippi State University.

    \item Demonstrating the possibility of promptly providing personalized, real-time information, thereby augmenting user engagement through the continuous availability of the chatbot. 

    \item Showcasing the application's effectiveness through rigorous evaluation, validating its performance and user satisfaction.

\end{itemize}

The rest of this article is divided into five connected sections. In Section \ref{background}, we present the background and related work. Following that, in Section \ref{methodology}, we explain the architecture and methodology. Section \ref{result-analysis} provides a detailed analysis of experimental results. Moving on to Section \ref{discussion}, we provide implementation details and discuss the limitations and future works. Finally, we conclude our paper.

\section{BACKGROUND AND RELATED WORK}
\label{background}

In this section, we briefly look into the pre-requisite background followed by exploring related research that focuses on development of chatbot applications in educational context. 

\subsection{Large Language Models (LLMs)}
Large Language Models (LLMs) like GPT-4, LLAMA3, and PaLM are at the forefront of computational linguistics, powered by Transformer-based  architectures \cite{vaswani2017attention} with vast parameter spaces, often exceeding hundreds of billions. These models rely on the self-attention mechanism within Transformers. LLMs excel in understanding and generating human language, reshaping the Natural Language Processing (NLP) landscape. They leverage various Transformer architectures and pre-training objectives, including decoder-only models (e.g., GPT2, GPT3), encoder-only models (e.g., BERT, RoBERTa), and encoder-decoder architectures like BART.

These architectures efficiently process sequential data, capturing intricate dependencies within text while enabling effective parallelization. LLMs integrate prompting or in-context learning, enhancing text generation by incorporating contextual information. This capability facilitates coherent and contextually relevant responses, fostering interactive question-and-answer engagements \cite{chang2024survey}.

\subsection{Retrieval Augmented Generation (RAG)}
Pre-trained Large Language Models (LLMs) are proficient at acquiring extensive knowledge but lack memory expansion or revision capabilities, leading to errors like hallucinations. To address this, hybrid approaches like Retrieval Augmented Generation (RAG) have emerged \cite{petroni2019language, ji2023survey, lewis2020retrieval}.

RAG integrates input sequences with information retrieved from corpus of an external data source, enriching context for sequence generation. The retriever component selects the top $k$ text passages relevant to the input query, augmenting the model's understanding and enhancing output sequence generation. This process is governed by the equation:  $p_n(z|x)$ where $p_n$ represents the retriever component with parameters $n$ (number of documents or passages a user wants to retrieve), selecting relevant passages $z$ from the knowledge database given input $x$.

\subsection{Related Works}

Recent research on educational chatbots explores various areas such as application fields, objectives, learning experiences, design approaches, technology, evaluation methods, and challenges. Studies have shown that educational chatbots are used in health advocacy, language learning, and self-advocacy. They can be flow-based or powered by AI, facilitating answering Frequently Asked Questions (FAQs), performing quizzes, recommending activities, and informing users about various events \cite{cunningham2019review}\cite{wollny2021we}. Chatbots have been found to improve students' learning experiences by motivating them, keeping them engaged, and providing immediate online assistance \cite{okonkwo2021chatbots}. Additionally, chatbots make education more accessible and available \cite{wollny2021we}. Design aspects such as the role and appearance of chatbots are significant factors that affect their effectiveness as educational tools \cite{martha2019design}. Chatbots are designed using various methods, including flow-based and AI-based approaches, and can incorporate speech recognition capabilities \cite{winkler2020sara}. Technologies used to implement chatbots include Dialogflow \footnote{https://cloud.google.com/dialogflow} and ChatFuel\footnote{https://chatfuel.com/} among others. These technologies impact chatbot performance and quality, necessitating careful selection during design and development \cite{perez2020rediscovering}. Flow-based chatbots, such as those powered by Dialogflow, can provide structured interactions based on predetermined scripts, while AI-based chatbots leverage machine learning and NLP to offer more flexible and dynamic interactions.



In regards to assessment of the effectiveness of educational chatbots, evaluation methods such as surveys, experiments, and evaluation studies are used, measuring acceptance, motivation, and usability \cite{hobert2019say}\cite{perez2020rediscovering}\cite{hwang2023review}. Surveys gather feedback from students and educators regarding their experiences with chatbots, while experiments may involve testing chatbots in controlled settings to measure their impact on learning outcomes. Evaluation studies provide deeper insights into how chatbots perform in various educational scenarios and how users perceive their usefulness. In terms of interaction styles, research examines whether chatbots are user-driven or chatbot-driven, depending on who controls the conversation \cite{winkler2020sara}\cite{cunningham2019review}. Chatbot-driven interactions often involve more automated and guided conversations, while user-driven interactions prioritize user input. 
Striking a balance between these approaches can result in more natural and effective communication. However, it's important to acknowledge that achieving this balance necessitates addressing substantive challenges to optimize the chatbot's applicability across diverse contexts, including the field of education.

Ethical considerations, such as compliance with educational norms and safeguarding user data, assume paramount importance \cite{okonkwo2021chatbots, tokayev2023ethical}. Leveraging novel methodologies in their development, we aim to navigate these issues more effectively. Moreover, we confront persistent programming complexities and the importance of sustaining chatbot utility amidst educational evolution \cite{adamopoulou2020overview, adamopoulou2020chatbots}. By harnessing advancements in technology, we endeavor to bolster our chatbots' resilience to these challenges. These collaborative endeavors offer a strategic direction, utilizing technological advancements to refine educational chatbots. Furthermore, the language model (conversational chatbot) contends with conceptual challenges essential for its operational efficacy, requiring careful research focus.

Insights from studies such as \cite{tripathi2023experimental} reveal how language models such as BERT establish relationships between expressions and queries, shedding light on chatbot interaction styles and response quality. This study contributes to understanding how advanced language models can be integrated into chatbots for more nuanced and context-aware responses. \cite{kuhail2023interacting} discusses the gap between chatbot responses and user intent, which can be more pronounced in complex university settings. Chatbots in academic environments often encounter questions that require a deep understanding of the subject matter and context. This necessitates the use of sophisticated models that can handle intricate queries and provide accurate and relevant responses. \cite{tripathi2023experimental, kuhail2023interacting} underscores the importance of understanding and controlling the context of language models, thereby guiding our efforts to integrate advanced language models into chatbots for more nuanced and context-aware responses. Their context-aware approach has been instrumental in shaping our chatbot's unique capabilities.


The integration of chatbots within university platforms and metaverses offers promising avenues for enhancing user experience and facilitating learning. For instance, \cite{xie2023chatbot} demonstrate how chatbots in metaverse-based university platforms offer instant, personalized support for tasks such as course navigation and answering FAQs, leveraging NLP and machine learning to streamline information dissemination and reduce administrative burdens. This kind of integration not only facilitates academic processes but also helps in addressing students' concerns promptly, ensuring smoother academic experiences. In specific university contexts, \cite{chandra2019indonesian} develops a question-answering system for an Indonesian university admissions using Sequence-to-sequence learning. This system demonstrates how chatbots can be employed in specialized areas to address particular challenges, such as providing guidance during the admissions process. Similarly, \cite{oliveira2023introducing} introduce a dynamic chatbot enhancing student interaction by covering admissions, academic assistance, and event information, prioritizing user feedback for accuracy, reliability, and safety. Frequent updates ensure that chatbots maintain relevance and continue to serve as effective tools for student support. Moreover, \cite{martinez2024designing} presents TutorBot+, which employs LLMs like ChatGPT to offer feedback in programming courses. Their quasi-experimental research shows positive impacts on students' computational reasoning abilities, illustrating the potential of such interventions in education. TutorBot+ demonstrates the benefits of integrating advanced AI models to support students in understanding complex programming concepts, potentially transforming how computational subjects are taught.

\section{\AlgoName Architecture \& Methodology}
\label{methodology}

\begin{figure*}[htb]
\centering
    \includegraphics[scale=0.99]{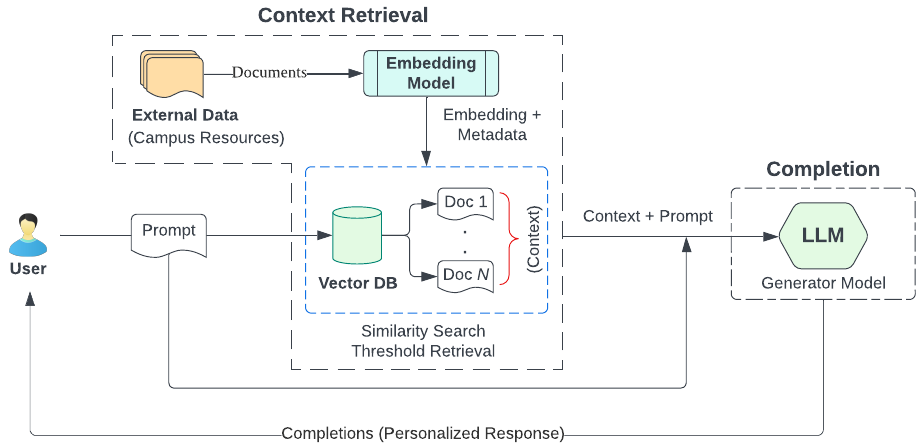}
    \caption{Overview of \AlgoName's two phase architecture. The first phase \emph{Context retrieval} is responsible to retrieve relevant documents based on the user prompt. The second phase,  \emph{Completion} responsible of generating personalized responses utlilizing retrieved documents as context along with user prompt.}\label{fig:arcchitecture}
\end{figure*}
This section describes the architecture of \AlgoName consisting two main phases: \emph{context retrieval} and \emph{completion} as shown in Fig. \ref{fig:arcchitecture}. The first phase retrieves documents relevant to the user prompt.  The second phase utilizes these retrieved documents and user prompts to generate contextual responses referred to as completions. The subsequent subsections will provide a comprehensive breakdown of each phase, discussing their functionalities and methodologies.

\begin{figure}[htb]
\centering
    \includegraphics[scale=1]{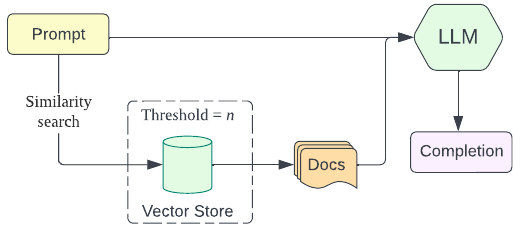}
    \caption{Similarity score threshold retrieval.}\label{fig:retriever}
\end{figure}

\subsection{Context Retrieval}
\label{context_retrieval}
Retrieval in \AlgoName involves obtaining pertinent information from an external data source to establish context for completions.
This phase takes a prompt (query) as an input and produces chunks of documents relevant to the prompt. In our context, the external data is the university resources available through Mississippi State University's Website \footnote{https://www.msstate.edu/}. We curate data of 42 different department within the university using web crawlers. These include \emph{academic departments, financial aid,  admissions, housing, dinning services, library, health center} etc. Inclusion of these campus resources as external data source is to ensure \AlgoName is comprehensive enough to answer diverse question. For example, a user might ask a question such as \emph{``What are the funding opportunities available for graduate students in the CSE department?"}. Followed by the question \emph{``Who do I contact if I have additional questions about majors or attending MSU?"} To answer the first question the system should have information about funding opportunities within CSE, whereas to answer the second question, information about \emph{academic counselors} should be present in the external data source. Please refer to Section \ref{dataset} for detailed explanation on how we curate these data, prepare and process for this phase.

The first step in this phase is to transform the external data source. This step relies on two important components: an \emph{embedding model} and a \emph{vector database}. Embeddings refers to functions that map or transforms raw input data to low-dimensional vector representations while retaining important semantic information about the inputs \cite{song2020information}. On the other hand, the vector database is a type of database that stores data as high-dimensional vectors that are usually generated by applying embedding functions to the raw data \cite{han2023comprehensive}, such as text in our case. It supports complex and unstructured data and allows fast and accurate similarity search and retrieval. \AlgoName utilizes an embedding model to vectorize the external data sources, in particular, we leverage a \emph{text-embedding-3-large} model managed through API calls. These vectors are subsequently stored in Chroma DB \cite{chroma} an in-memory vector database.

For efficient context retrieval process we use vector store-backed retriever technique provided by LangChain \cite{langchain}. 
It utilizes vector store to retrieve documents. In general the vector store retriever uses two types of search methods including Maximum Marginal Relevancy (MMR) and Similarity Search. In this work, we have leveraged \emph{similarity score threshold retrieval}, depicted in Fig.\ref{fig:retriever} as our searching mechanism. This searching strategy returns all the possible results to a user prompt based on the minimum similarity threshold we define. The output of this phase is the relevant documents that serves as context for subsequent completions phase. 
\subsection{Completion}
The second phase is the \emph{completion} which is also referred to as \emph{response generation}. We utilize a gpt based LLM for completions, in particular, we leverage OpenAI's \emph{gpt-3.5-turbo} as our base generator model. The input to the generator consists of retrieved document chunks and the user prompt. Then, the generator model, or LLM, uses this information as a guideline to produce acurate and relevant responses (completions). Response generation in \AlgoName is managed through OpenAI's API calls.

An example of completion for the user prompts \emph{"What are the funding opportunities available for graduate students in the CSE department?"} and \emph{"Who do I contact if I have questions about attending MSU} can be seen in Fig. \ref{box_final_Completions}. In the first example, a user seeks information about funding opportunities within the CSE department for graduate studies. The completion, in this case, is a contextualized response generated by combining relevant information retrieved from campus resources, such as the academic department (CSE), with the user prompt. In the second example, the user is seeking information about admission to Mississippi State University. To answer this, the system must possess knowledge about campus resources, such as admissions and scholarships. As mentioned earlier, \AlgoName not only has knowledge about academic departments but also other campus resources, including admissions and scholarships. By utilizing admissions and scholarships as external data sources, the system returns the relevant document reflecting information about admissions and scholarships, along with contact details, as a completion for the user prompt.
\begin{center}
\begin{minipage}{8.8cm}
        
        \begin{tcolorbox}[enhanced,attach boxed title to top center={yshift=-1mm,yshifttext=-1mm},
            colback=gray!10!white,colframe=gray!90!black,colbacktitle=gray!80!black, left=0.1mm, right=0.5mm, boxrule=0.50pt]
            \scriptsize
           {\fontfamily{qcr}\selectfont

            \textbf{Prompt}: What are the \textcolor{red}{funding opportunities} available for \textcolor{red}{graduate students} in the \textcolor{red}{CSE department}?\\ \\
            
                \textbf{Completion}: \textcolor{red}{Graduate students} in the \textcolor{red}{Computer Science and Engineering (CSE)} department at Mississippi State University have access...funding opportunities...\textcolor{red}{graduate assistantships, research grants, fellowships, scholarships, and industry partnerships}...research grants cover project expenses. \\ \\
            
                \textbf{Prompt}: Who do I \textcolor{red}{contact} if I have questions about attending MSU? \\ \\
                \textbf{Completion}: For queries on admissions at Mississippi State University, reach out to the \textcolor{red}{Office of Admissions and Scholarships} at \textcolor{red}{662-325-2224} or \textcolor{red}{admissions@msstate.edu}.
           }

        \end{tcolorbox}
       \vspace{-14pt}
        \noindent\begin{minipage}{\textwidth}

\end{minipage}
\end{minipage}
\end{center}
\captionof{figure}{A graphical depiction of response generation. Red highlights indicate \AlgoName's ability to contextualize and generate factually grounded response.}\label{box_final_Completions}

\section{Experiments and Results}
\label{result-analysis}

To evaluate the performance of our system, we adopt a two-fold approach including both \emph{quantitative} and \emph{usability assessment} methods. For the quantitative evaluation (See Section \ref{app: data-driven}), we utilize the RAGAS \cite{es2023ragas} framework, while the SUS is adopted for usability assessment (See Section \ref{app: qualitative-assessment}). In the following subsection, we first discuss the dataset and steps we took to prepossess them and then provide a detailed explanation of our evaluation approaches.

\subsection{Dataset Description \& Preparation}
\label{dataset}


To ensure a comprehensive chatbot system capable of answering diverse questions—whether academic or non-academic—we initially developed a web scraper to gather information on various campus resources at Mississippi State University. This collection would then serve as an external data source in our pipeline. We scraped various campus resources including academic departments, financial aid, scholarships, housing, dining, parking, and police. In total, we scraped 42 campus resources into a JSON file. Each JSON file includes the following information: the URL, title, and content of the scraped webpage, all wrapped into a JSON object. 
We consolidated the individual files into a master JSON file which serves as an external data source and is ingested into our RAG pipeline. A subset of the data utilized by \AlgoName can be observed in Table \ref{data}. 

\quad To enhance retrieval accuracy, we first preprocess the JSON file. This preprocessing step involves removing noise, such as undesirable Unicode characters, redundant, and unnecessary information. We then implement a recursive chunking strategy, with a chunk size of 8000  and an overlap of 1200 characters. 
This step is 
crucial for optimizing the performance of RAG chatbot systems  
with the objective 
of ensuring that our 
chatbot generates 
an accurate response that is contextually appropriate. Subsequently, we transformed the textual data into vectorized representations utilizing an \emph{embedding model} (Refer to Section \ref{context_retrieval} to learn for more details on embedding models.).










\begin{table}[]
\centering
\caption{A subset of an external data source containing campus resources, including both academic and non-academic departments, indicating the total number of tokens associated with each.}
{\renewcommand{\arraystretch}{1.30}
\begin{tabular}{c|l|l}
\cline{2-3}
\multicolumn{1}{l}{}                                    & \textbf{Departments}                    & \textbf{\# of Tokens} \\ \cline{2-3}
\multicolumn{1}{c}{\multirow{11}{*}{\rotatebox[origin=c]{90}{Campus Resources}}} & Computer Science and   Engineering      & 200623                      \\ \cline{2-3} 
\multicolumn{1}{c}{}                                   & Chemical Engineering                    & 118271             \\ \cline{2-3} 
\multicolumn{1}{c}{}                                   & Electrical and Computer Engineering     & 328558                      \\ \cline{2-3} 
\multicolumn{1}{c}{}                                   & Industrial and Systems Engineering      & 22390                      \\ \cline{2-3} 
\multicolumn{1}{c}{}                                   & Agricultural and Biological Engineering & 79978                     \\ \cline{2-3} 
\multicolumn{1}{c}{}                                   & Civil and Environmental Engineering     & 61071                       \\ \cline{2-3} 
\multicolumn{1}{c}{}                                   & Aerospace Engineering                   & 37812                       \\ \cline{2-3} 
\multicolumn{1}{c}{}                                   & Biomedical Engineering                  & 256761                   \\ \cline{2-3} 
\multicolumn{1}{c}{}                                   & Housing                                 & 132193                    \\ \cline{2-3} 
\multicolumn{1}{c}{}                                   & Admission                               & 276972                      \\ \cline{2-3} 
\multicolumn{1}{c}{}                                   & MSU Police                               & 16629                      \\ \cline{2-3} 
\end{tabular} }

\label{data}
\vspace{-5mm}
\end{table}

\subsection{Quantitative Evaluation}
\label{app: data-driven}
To evaluate \AlgoName's ability to produce contextually appropriate responses, we utilize the RAGAS framework \cite{es2023ragas}. We choose this framework because it is specifically designed to assess RAG pipelines. Other popular evaluation metrics such as ROUGE \cite{lin2004rouge} and BLEU \cite{papineni2002bleu} are not suitable in our context. This is because ROUGE is generally used to evaluate summarization tasks, while BLEU is designed to evaluate language translation tasks.

\begin{table*}[ht!]
\centering
\caption{Overview of results: Retrieval scores pertain to the \emph{context retrieval} phase of the architecture, where \emph{prec.} refers to context precision, and recall refers to context recall. Generation scores pertain to the \emph{completion phase}, where \emph{faith} stands for faithfulness and \emph{rel.} for answer relevancy. The end-to-end evaluation showcases \AlgoName's efficiency in generating contextually relevant and accurate answers through metrics such as answer similarity and answer correctness.}
\label{tab:different_retrieval_scores}
\begin{tabular}{@{}lcccccccc@{}}
\toprule
& \multicolumn{2}{c}{Retrieval} & \multicolumn{2}{c}{Generation} & \multicolumn{1}{c}{RAGAS Score} & \multicolumn{2}{c}{End-to-End Evaluation} \\
\cmidrule(r){2-3} \cmidrule(r){4-5} \cmidrule(r){6-6} \cmidrule(r){7-8}
Category & Prec. & Recall & Faith. & Rel. & Harmonic Mean & Answer Similarity & Answer Correctness \\
\midrule
Engineering Programs & 0.98 & 0.96 & 0.99 & 0.97 & 0.97 & 0.8434 & 0.8620 \\
General Inquiry & 0.95 & 0.97 & 0.98 & 0.96 & 0.96 & 0.7764 & 0.8123 \\
Research Opportunities & 0.97 & 0.98 & 0.96 & 0.99 & 0.97 & 0.8245 & 0.8841 \\
University Resources & 0.96 & 0.99 & 0.97 & 0.98 & 0.97 & 0.8317 & 0.8923 \\
\bottomrule
\end{tabular}
\vspace{-4mm}
\end{table*}

\quad We evaluate both phase of \AlgoName architecture (See section \ref{methodology}) i.e. \emph{context retrieval} and \emph{completion}. To evaluate the retrieval, we employ two metrics such as \emph{context precision and context recall}. The first metric represents the Signal-to-Noise Ratio (SNR) of retrieved context, while the second metric evaluates whether the retriever has the ability to retrieve all the relevant evidence to answer a question. Similarly, to evalaute \emph{completion or generation} we employ \emph{faithfullness} and \emph{answer relevance} metrics. Faithfulness evaluates how factually accurate the generated answer is while answer relevance evaluates how relevant the generated answer is to the question. The final RAGAS score, representing the harmonic mean of these four metrics, falls within a range of 0 to 1, with 1 denoting optimal generation. This score serves as a singular measure of a QA system's performance. Therefore, the RAGAS score is essential for assessing the overall performance and relevance of \AlgoName in its targeted educational environments.


\quad To 
conduct phase wise evaluation, we first crafted a set of questions and their ground truth pertaining to \emph{engineering programs, general inquiries, research opportunities}, and \emph{other university resources}. We report RAGAS score of 0.97, 0.96, 0.97 and 0.97 for these categories respectively in Table \ref{tab:different_retrieval_scores}. These score underlines both retrieval and completion component are efficient.

\quad We also conduct end-to-end evaluation to measure overall performance of \AlgoName, as it directly affects the user experience. Metrics such as \emph{answer similarity} and \emph{answer correctness} are employed to assess the overall performance, ensuring a comprehensive evaluation. In particular, \emph{answer similarity} scores that reflect strong alignment with ideal responses are reported to be high in cases when questions about engineering programs and research opportunities are asked, with scores of 0.8434 and 0.8317 respectively. Moreover, \emph{answer correctness}, which indicates high factual accuracy, is reported to be high when the system is asked questions about university resources and research opportunities, at 0.8923 and 0.8841 respectively. Overall, these metrics suggest that \AlgoName effectively retrieves relevant and accurate answers.

\subsection{Usability Assessment}
\label{app: qualitative-assessment}
\begin{figure*}[ht!]
\centering
\includegraphics[width=1\textwidth, height=0.34\textheight]{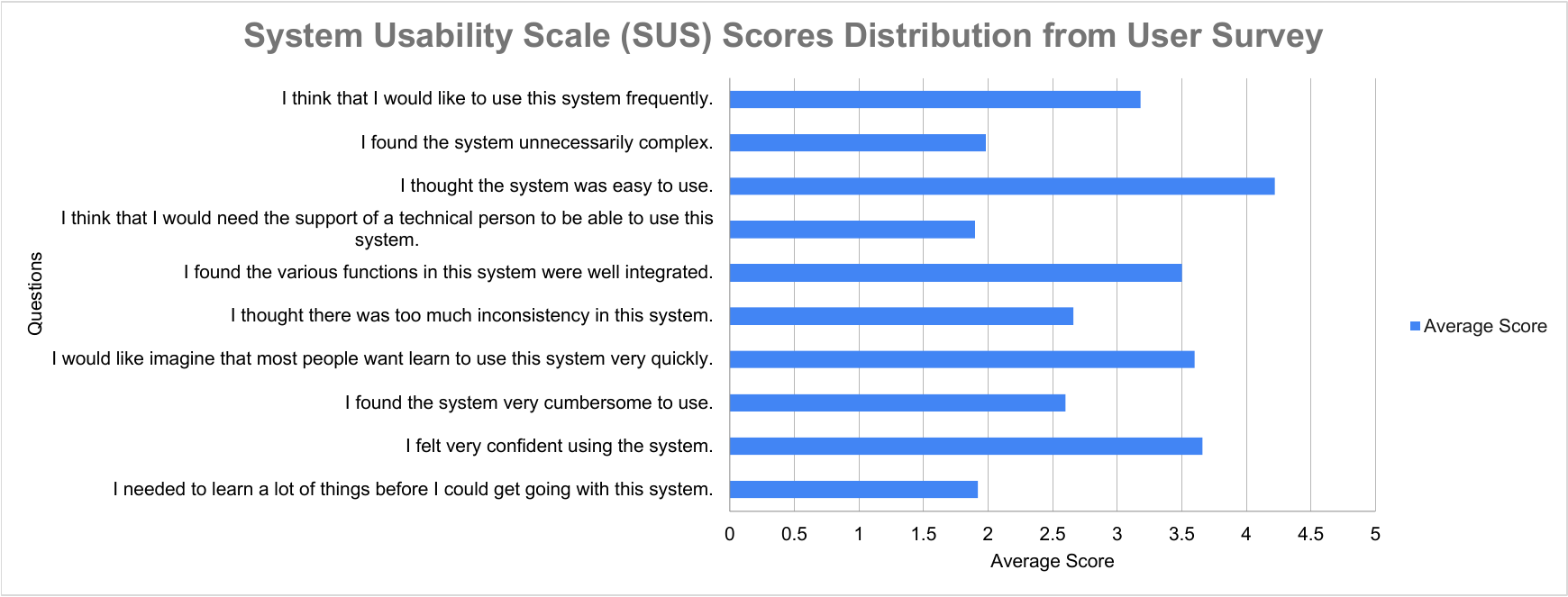} 


\caption{Distribution of average System Usability Scale (SUS) scores.}

\label{fig:bar-sus}
\end{figure*}

To further understand the user experience when using \AlgoName, we perform a subjective satisfaction survey using the System Usability Scale (SUS) \cite{vlachogianni2022perceived} - a widely reliable method that accesses systems usability through set of questionnaire. Given the expensive nature of this evaluation we engage a panel of 50 graduate and undergraduate students undertaking CSE8011 (Seminar course) at Mississippi State University. The participants were tasked to answer set of 10 questions as depicted in \ref{fig:bar-sus}, each offering  five response options ranging from “strongly agree” to “strongly disagree”. We then collected their feedback and calculated an average SUS score of 67.75. The feedback results indicated satisfactory usability with a room for improvement for future iterations of our system. 




\section{Detailed Analysis and Insights}
\label{discussion}
In this section, we discuss 
implementation details, where we explain the technical process behind developing \AlgoName. Then, we discuss constraints and shortcomings encountered, and provide our plan for the future.


\subsection{Implementation Detail}
\label{implementation-detail}
For data curation we employed a multi-thread web crawler with the Scrapy Python library to collect data from over 42 campus resources (See Section \ref{dataset} for details). We carefully selected important HTML div tags that comprised of relevant information about a topic. This process was semi-automatic in nature because every HTML pages were differently formatted with different div ids. Manual div selection also allowed us to remove noise to some extent. The data was exported to JSON file format with url, topic and content. Individual JSON files for each of the campus resources was then consolidated into a master JSON file for comprehensive retrieval. 

\quad We predominantly use LangChain framework to develop \AlgoName. First, we preprocess master JSON into smaller chunks using Recursive Character Text Splitter splitting strategy. Given the nature of our data we opted for 8000 chunk size with 1200 overlap. We then apply an embedding function on these chunks utilizing OpenAI's \emph{text-embedding-3-large} model and store the vectors in Chroma DB. This step allowed us to retrieve documents relevant to specific user prompts. In our case, we utilize \emph{vectorstore} for \emph{context retrieval} with a similarity search threshold as our search strategy (See Section \ref{context_retrieval} for more details). For completion or response generation we leverage OpenAI's \emph{gpt-3.5-turbo} model. Both embedding and response generation is managed through API calls. 

\quad \AlgoName is built with Django framework using python. For front-end we utilize HTML, CSS and Javascript. The current version of our system has not only question-answering functionality but also user sign up and log in feature. Once a user is registered they can ask queries, they can see previous conversations, delete conversations, and email conversations. Our application is deployed through a third-party cloud service for accessibility.


\subsection{Limitations \& Future Direction}
\label{limitations}

Despite the achievements in developing our educational chatbot, several significant challenges currently limit its broader application.\AlgoName does not currently have Automatic Speech Recognition (ASR) capability, which might hinder its use among visually impaired, disabled, or elderly users. Additionally, given that Mississippi State University hosts a number of international students annually from non-English speaking countries, it currently lacks multi-lingual support. In terms of technical limitations, our retrieval system sometimes fails to provide accurate or relevant results, occasionally producing incorrect information, a phenomenon known as `hallucinations'. We are also limited by a maximum number of output tokens, which is 4096, and a context window of 16k. This sometimes hinders system's ability to capture the full length of the conversation in the memory buffer.

\quad To address the limitations discussed above and enhance \AlgoName's functionality and usability, we are planning several key upgrades. These include adding support for multiple languages to cater to a diverse user base, integrating ASR and text conversion features to enable various interaction modes, and improving the retrieval algorithms to boost the accuracy and relevance of the information provided. Moreover, in response to the token limitations of the OpenAI API, we aim to apply the map-reduced document chain approach from LangChain. Through these improvements, we aim to transform \AlgoName into a more reliable and accessible educational tool.

\section{Conclusion}
\label{conclusion}

This study highlights the significant potential of AI-based chat systems in improving communication and access to information regarding university resources. Our system, \AlgoName integrates large amounts of university data, including academic programs, campus facilities, student service as external data corpus into its RAG pipelines for domain-specific question and answering tasks. By incorporating this external data corpus, our system ensures the delivery of precise and contextually relevant responses to both academic and non-academic user inquiries. The comprehensive end-to-end evaluation process demonstrated \AlgoName's efficiency in generating contextually relevant and accurate answers as measured by metrics such as answer similarity and correctness. Furthermore, system usability experiments employing the SUS indicated that \AlgoName is practical and effective for real-world usage, affirming its reliability and the positive user experience it offers. The positive outcomes of using \AlgoName at Mississippi State University suggest promising opportunities for broader implementation. This system could be adapted for use in other universities or different sectors and can be viewed as enterprise document retrieval systems that enhance user engagement and information access.

\section*{Declaration of Competing Interest}

The authors declare that the research was conducted in the
absence of any commercial or financial relationships that could
be construed as a potential conflict of interest.

\section*{Acknowledgement}

This work was supported by the PATENT Lab (Predictive
Analytics and Technology Integration Laboratory) at the
Department of Computer Science and Engineering, Mississippi
State University.

\bibliographystyle{IEEEtran}
\bibliography{references}

\end{document}